\documentclass{emulateapj}

\usepackage{amssymb}
\usepackage{amsmath}
\usepackage{graphicx}

\newcommand{\kapGP}{\kappa_f}

\newcommand{\kGP} {k_f}
\newcommand{\vGP} {v_f}
\newcommand{\kP}  {k_P}
\newcommand{\lAD} {\lambda_{AD}}
\newcommand{\lO}  {\lambda_{\Omega}}
\newcommand{\led}  {\lambda_{e}}
\newcommand{\lnum}  {\lambda_{num}}
\newcommand{\tAD} {\tau_{AD}}

\newcommand{\tGP} {\tau_f}
\newcommand{\cAi} {c_{Ai}}

\newcommand{\RM} {R_{M}}
\newcommand{\RAD} {R_{AD}}
\newcommand{\mbfB}{\mathbf{B}}
\newcommand{\mbfu}{\mathbf{u}}
\newcommand{\mbfui}{\mathbf{u}_i}
\newcommand{\mbfun}{\mathbf{u}_n}
\newcommand{\mbfuD}{\mathbf{u}_D}
\newcommand{\mbfuGP}{\mathbf{u}_{GP}}
\newcommand{\mbfz}{\mathbf{z}}
\newcommand{\mbfn}{\mathbf{n}}
\newcommand{\mbfk}{\mathbf{k}}
\newcommand{\mbfx}{\mathbf{x}}
\newcommand{\nuin}{\nu_{in}}
\newcommand{\mbfnabla}{\mathbf{\nabla}}
\newcommand{\f}   {\frac}

\newcommand{\ddt}{\frac{\partial}{\partial t}}
\newcommand{\mq}{\langle q\rangle}
\newcommand{\dq}{\delta q}
\begin{document}

\title{Turbulent Ambipolar Diffusion: Numerical Studies in 2D}

\author{Fabian Heitsch\altaffilmark{1,2}}
\author{Ellen G. Zweibel\altaffilmark{1,2,3}}
\author{Adrianne D. Slyz\altaffilmark{4}}
\author{Julien E.G. Devriendt\altaffilmark{4}}
\altaffiltext{1}{U Wisconsin-Madison, 475 N Charter St, Madison,
                 WI 53706, U.S.A.}
\altaffiltext{2}{JILA/U Colorado, Boulder, CO 80309-0440, U.S.A.}
\altaffiltext{3}{Center for Magnetic Self-Organization in Laboratory \& Astrophysical
Plasmas}
\altaffiltext{4}{U Oxford, Keble Road, OX1 3RH Oxford, UK}

\lefthead{Heitsch et al.}
\righthead{Turbulent Ambipolar Diffusion}

\begin{abstract}
Under ideal MHD conditions the magnetic field strength should be
correlated with density in the interstellar medium (ISM). However,
observations indicate that this correlation is weak.
Ambipolar diffusion can decrease the flux-to-mass ratio in weakly
ionized media; however, it is generally thought to be too slow to
play a significant role in the ISM except in the densest molecular
clouds. Turbulence is often invoked in astrophysical problems to
increase transport rates above the (very slow) laminar values predicted
by kinetic theory. We
describe a series of numerical experiments addressing the problem of
turbulent transport of magnetic fields in weakly ionized gases. We
show, subject to various geometrical and physical restrictions, that turbulence in
a weakly ionized medium rapidly diffuses the magnetic flux to mass ratio $B/\rho$
through the buildup of appreciable ion-neutral drifts on small scales. These
results are applicable to the fieldstrength - density correlation in the ISM,
as well as the merging of flux systems such as protostar and accretion disk fields or
protostellar jets with ambient matter, and the vertical transport of galactic
magnetic fields.
\end{abstract}
\keywords{diffusion --- MHD --- turbulence --- methods:numerical 
          --- ISM:magnetic fields}

%
%
\section{Introduction}\label{s:motivation}

Two dimensionless parameters control the
degree to which galactic magnetic fields are frozen to the interstellar gas.
One, the magnetic Reynolds number $\RM$, is the ratio of the Ohmic
diffusion time to the dynamical time, and is typically of order 10$^{15}$ - 10$^{21}$.
The second, the ambipolar Reynolds number $\RAD$, is the ratio of the ion-neutral 
drift time to the dynamical time. This number is typically many orders of magnitude 
less than $\RM$, and can approach unity in dense molecular gas. Based on these 
estimates, magnetic fields should be nearly perfectly frozen to the plasma 
component of the gas, and generally quite well frozen to the neutrals, except
in the densest, nearly neutral regions.

Thus, the ratio of magnetic fieldstrength to gas density $B/\rho$ is determined
primarily by dynamical rather than by microscopic processes. Parameterizing 
the $B-\rho$ relation by $\kappa\equiv d\ln{B}/d\ln{\rho}$, one finds $\kappa=1$ 
for compression perpendicular to $\mbfB$, $\kappa=2/3$ for isotropic compression, 
$\kappa\sim 1/2$ for self-gravitating, magnetically subcritical clouds, and 
$\kappa=0$ for compression parallel to $\mbfB$.

Observations of the  $B-\rho$ relation in molecular gas indeed show that the 
strongest fields are associated with the densest gas (\citet{CRU1999}, 
\citet{BMR2001}, \citet{STC2002}). The correlation is consistent with $\kappa
\sim 0.5$, although there is so much dispersion, particularly when upper limits 
are included, that this relation is perhaps only an upper envelope. In atomic gas, the  
$B-\rho$ relation is consistent with $\kappa\sim 0$ over three orders of magnitude 
in $\rho$ \citep{TRH1986}. Observational effects alone can introduce substantial
scatter because all measurements are averages along the line of sight and over the 
telescope beam width, $\rho$ may not be accurately determined, and because only the 
line-of-sight component of $\mbfB$ is measurable. However, even allowing for the 
possibility that the observational scatter is large,  
the $B-\rho$ relation is strikingly flat. Moreover, there is a mean 
density jump of at least a factor of $50$ when going from atomic to molecular 
gas, whereas the corresponding mean magnetic field strength seems to increase by a 
factor of two or three at most.

At first sight, the simplest explanation of the observed  $B-\rho$ relation is 
that dense regions in the ISM arise primarily from compression parallel to 
$\mbfB$. Arguments against this as the sole explanation are quantitative rather than 
qualitative. In order to collimate the flow, the magnetic energy density should 
dominate the turbulent energy density, but the field is at or below equipartition. 
And, if giant molecular clouds are assembled by 1D compression, they must
sweep up material over nearly a kpc, too large a scale on which to expect coherent 
flow \citep{MES1985}.

In astrophysical environments, the microscopic diffusivities - whether viscous, 
resistive, or chemical - are often far too small to explain the transport that 
apparently takes place. However, diffusion rates can 
be enhanced by turbulence. Turbulence accelerates transport because it creates 
small scale structure, which diffuses faster than the original large scale 
structure and smoothes the large scale gradients. In a turbulent flow with 
characteristic velocity $u$ and correlation time $\tau$, the effective
diffusivity $\led$ is argued to be of order $u^2\tau$ (see eq. [\ref{e:led}]).

Therefore, in this paper, we investigate whether turbulence has
a similar effect on the transport of magnetic fields with respect to the neutral gas.
Because the interstellar magnetic field is subject to ion-neutral drift at scales
much larger than the resistive scale, we ignore resistive effects, except to control 
numerical diffusion in our calculation, and concentrate on the properties of ion-neutral
drift in a turbulent medium. Although our work is motivated by the $B - \rho$ 
relation observed in the interstellar medium, it is also relevant to other problems, 
including transport of magnetic fields in weakly ionized accretion disks and 
entrainment of molecular gas by protostellar outflows.

In \S\ref{s:background}, we review the theoretical basis for turbulent
ambipolar drift and summarize previous results. In \S\ref{s:description} we
introduce our model. Section \ref{s:numerics} is a description of the numerical
method and its validation. The main results are presented in \S\ref{s:results}.
Finally, we mention other applications in the concluding section,
together with a summary.

%
%
\section{Background\label{s:background}}

Consider a medium in which the ionization is sufficiently low that the neutral
and total densities, $\rho_n$ and $\rho$, are interchangeable, as are the neutral
and center of mass velocities $\mbfun$ and $\mbfu$. Under these conditions, 
the continuity equation
\begin{equation}\label{e:continuity}
\f{\partial\rho}{\partial t}=-\mbfnabla\cdot\rho\mbfu
\end{equation}
and magnetic
induction equation
\begin{equation}\label{e:induction}
\f{\partial\mbfB}{\partial t} = \mbfnabla\times\left(\mbfui\times\mbfB\right)
\end{equation}
can be combined to yield an equation for $\mbfB/\rho$
\begin{equation}\label{e:genBrho}
\left(\f{\partial}{\partial t}+\mbfu\cdot\mbfnabla\right)\f{\mbfB}{\rho}=
\f{\mbfB}{\rho}\cdot\mbfnabla\mbfu
+\f{1}{\rho}\mbfnabla\times\left(\mbfuD\times\mbfB\right),
\end{equation}
where $\mbfuD\equiv\mbfui-\mbfun$ is the relative drift between the ions and the
neutrals. The first term on the right hand side of equation~(\ref{e:genBrho})
represents fieldline stretching, while the second term represents ambipolar
diffusion.

In this paper, we restrict ourselves to a geometry in which all velocities are in
the ($x,y$) plane, $\mbfB=\hat\mbfz B$, and all
quantities are independent of $z$. Under these restrictions there is no
fieldline stretching, and equation~(\ref{e:genBrho}) reduces to
\begin{equation}\label{e:2DBrho}
\left(\f{\partial}{\partial t}+\mbfu\cdot\mbfnabla\right)\f{B}{\rho}=-\f{1}{\rho}
\mbfnabla\cdot(\mbfuD B).
\end{equation}

If the timescales of interest are long compared to the ion-neutral collision time
$\nuin^{-1}$, $\mbfuD$ is determined by balancing the Lorentz force on the ions
against the frictional force on the neutrals (the so-called strong coupling
approximation). For the geometry assumed here, the Lorentz force
is due only to the magnetic pressure gradient, and
\begin{equation}\label{e:sca}
\mbfuD=-\f{1}{\rho_i\nuin}\mbfnabla\f{B^2}{8\pi}.
\end{equation}
Substituting equation~(\ref{e:sca}) into equation~(\ref{e:2DBrho}) yields
\begin{equation}\label{e:Brhodiff}
\left(\f{\partial}{\partial t}+\mbfu\cdot\mbfnabla\right)\f{B}{\rho}=\f{1}{\rho}
\mbfnabla\cdot\lAD\mbfnabla B,
\end{equation}
where
\begin{equation}\label{e:lAD}
\lAD\equiv \f{B^2}{4\pi\rho_i\nuin}
\end{equation}
is the ambipolar diffusivity.

Equation~(\ref{e:Brhodiff}) is almost, but not quite, an advection - diffusion equation
for $B/\rho$. The difference is that the diffusive flux is proportional to 
$\mbfnabla B$, not $\mbfnabla (B/\rho$).
As an immediate consequence,  $B/\rho$ will develop a large dispersion if $\rho$ 
itself is advected by the turbulence as a passive scalar (see \S\ref{ss:diffbrho}).

Equation~(\ref{e:Brhodiff}) predicts a characteristic ambipolar diffusion time 
$\tAD$ for a field with characteristic lengthscale $L_B$
\begin{equation}
  \tau_{AD}\equiv\frac{L_B^2}{\lambda_{AD}},
  \label{e:tad}
\end{equation}
We want to know for which turbulent flow speeds $u$ and length scales $L_e$ would such a 
field diffuse at a rate higher than the laminar rate. If the turbulence has a correlation 
time $\tau_c=L_e/u$, the turbulent diffusivity $\led$ is of order $u^2\tau_c=L_e u$ 
(see eq.~[\ref{e:led}]). The ratio of the ambipolar diffusion time $\tau_{AD}$  
to the turbulent diffusion time $\tau_t$ is just the ratio of the two diffusivities: 
$\tau_{AD}/\tau_t=\led/\lAD$. Therefore we are interested in the case $\led/\lAD > 1$. 
This condition will be satisfied if the magnetic field is well frozen to the turbulent 
eddies. The degree of freezing is measured by the eddy
ambipolar Reynolds number $\RAD(L_e,u)$, the ratio of the ambipolar diffusion time across
the eddy, $\tau_{AD}(L_e)$ to the eddy correlation time. Assuming the eddy size
$L_e$ is related to $u$ and $\tau_c$ by $L_e = u\tau_c$ we have
\begin{equation}\label{e:RAD}
\RAD(L_e,u)\equiv
\f{\tau_{AD}(L_e)}{\tau_c}=\f{L_eu}{\lAD}=\left(\f{u}{\cAi}\right)^2\tau_{c}\nuin.
\end{equation}
Inserting numerical values,
\begin{equation}\label{e:RADnum}
  \RAD(L_e,u)= 9.4\times10^{-9}\,L_e\,u
  \left(\frac{n_n}{B}\right)^2\frac{\mu_i\,\mu_n}{\mu_i+\mu_n}\,
  x_i,
\end{equation}
where $L_e$ is expressed in parsecs, $u$ in km s$^{-1}$, $B$ in Gauss, and $n_n$ in
cm$^{-3}$. The $\mu$ represent molecular weights, and $x_i$ is the ionization
fraction. The field is frozen to turbulent eddies which are larger than the size at
which $\RAD(L_e,u)=1$. For example, in
gas with an ionization fraction $10^{-3}$, $\mu_i/\mu_n\gg 1$, magnetic field
$B=5\mu G$, an internal velocity dispersion $u=1$, and neutral density $n_n=50$, this
critical lengthscale is about 10$^{-3}$pc, corresponding to a column density of about
$1.5\times 10^{17}$ cm$^{-2}$.

What diffusion rate is actually required to break the flux freezing and produce a flat
$B - \rho$ relation? Suppose a coherent density structure of size $L$ forms on a timescale
$\tau$, with an associated velocity $U$. The $B - \rho$ relation breaks down if the 
diffusion time is less than the formation time. In other words, if the relation
\begin{equation}\label{e:enhancement}
\lAD < LU < \led
\end{equation}
holds, then laminar ambipolar drift is too slow to break flux freezing but turbulence is
fast enough.

If $L$, $U$, $L_e$, and $u$ are related through the usual scaling laws for a turbulent 
cascade, then $U/u > 1$ whenever $L/L_e > 1$. Hence turbulent mixing can only destroy 
the $B-\rho$ relation in large scale structures which are controlled by additional physics, 
such as cooling or self gravity, or in regions with strong sources of small scale 
turbulence, such as velocity shear layers.

Equation~(\ref{e:Brhodiff}) is the basis of three recent studies of the
effects of turbulence on the rate of ambipolar drift. \citet{ZWE2002}
demonstrated accelerated diffusion at approximately the eddy rate $\led$ by finding
exact solutions of equation~(\ref{e:Brhodiff}) for random sequences of
incompressible stagnation point flows, valid for
a particular initial magnetic profile and initially constant density.
\citet{FAD2002} concentrated on the role of fluctuations in $B$, and hence
$\lAD$, in reducing $\tAD$. They showed that fluctuations in fieldstrength lead
to a corresponding dispersion in ambipolar diffusion times.
\citet{KID2002} analyzed equation~(\ref{e:Brhodiff}), assuming
$\mbfnabla\cdot\mbfu\equiv 0$ and constant $\rho$, under the standard assumptions
of quasilinear diffusion theory (QDT, see Appendix). 
They argued that the canonical turbulent diffusion rate $\led\sim u^2\tau$ is an upper 
limit which is approached for turbulence well frozen to the eddies.

Within the framework of QDT, the novel aspect of equation~(\ref{e:Brhodiff}) is the 
right hand side, which consists of a nonlinear diffusion operator applied to $B$, not 
$B/\rho$. The latter distinction is unimportant as long as $\rho$ is uniform. 
\citet{KID2002} argue that diffusion reduces turbulent transport, because in a highly 
diffusive system, the field is not advected by the flow. While 
equation~(\ref{e:genBrho}) shows that in the limit $\lambda_{AD} \equiv 0$ there is 
no diffusion in an absolute sense, we will see that the large scale component can decay by 
transferring power to the small scales.

Taken together, \citet{ZWE2002}, \citet{FAD2002}, and \citet{KID2002}
support the proposition that ambipolar drift is accelerated in a turbulent
medium, provided that the strong coupling approximation holds, and in
the 2.5D geometry assumed in all three papers. The diffusion rate is enhanced by
the development of small scale structure in the field, which increases the local
drift velocity, and by the growth of fluctuations in fieldstrength, which
increases the local diffusivity. When the former dominates, the diffusion rate
is close to the canonical turbulent value $u^2\tau$, and nearly independent of the
microscopic diffusivity.

Although these papers are suggestive, the picture they present is still incomplete. 
Due to the choice of initial condition, the difference between diffusion of $B$ and 
diffusion of $B/\rho$ is not addressed by any of the work, although the former 
describes transport of $B$ from an Eulerian volume and the latter from a Lagrangian 
one. The respective roles of irreversible flux transport by ion-neutral drift and 
breakup of large scale structure into small scale fluctuations is not discussed by 
the QDT calculation, while the stagnation point calculation has yet to be embedded 
in a globally valid flow model. In the remainder of this paper, we report on a series 
of numerical experiments of the same basic problem which address some, but not all, 
of these limitations.

%
%
\section{Description of the Problem\label{s:description}}

\subsection{Assumptions and Equations\label{ss:assumptions}}

As in \S\ref{s:background}, we consider the action of  2D flow (in the ($x,y$) 
plane) on a perpendicular magnetic field $\hat\mbfz B$. Because we anticipate the 
formation of structure on small scales, we do not assume strong coupling 
(eq.~[\ref{e:sca}]). Instead, we compute $\mbfui$ by solving the ion equation of motion
\begin{equation}
  \rho_i\left(\ddt\mbfui + (\mbfui\cdot\mbfnabla)\mbfui\right)
  = -\f{1}{8\pi}\mbfnabla B^2-\rho_i\nuin(\mbfui-\mbfun).
  \label{e:momentum}
\end{equation}

We assume that $\rho_i$ is related to $\rho_n$ through the condition of ionization
equilibrium. Although this breaks down on short timescales, we have argued elsewhere 
that it is generally an excellent approximation \citep{HEZ2003}. We also neglect ion 
pressure relative to magnetic pressure. This is a good approximation in weakly 
ionized interstellar gas except near magnetic nulls, which are precluded by the 
choice of initial conditions (see \S\ref{ss:ic}).

We simplify the problem further by treating $\mbfun$ as given. Although this so-called
kinematic approach is frequently made in turbulent mixing problems, it is not 
particularly accurate in the interstellar medium, because the magnetic and turbulent 
pressures are comparable. However, in weakly ionized gas there is a lengthscale below 
which the neutrals and ions are not well coupled (see \S\ref{sss:dispersionrelation}), 
and one can regard the dynamics as taking place below this critical scale.
One consequence of treating the neutrals kinematically is that $\rho$ itself behaves 
as a passive scalar. We could reach the same result by assuming zero temperature, 
and neglecting ion pressure.

With $\mbfui$ determined from equation~(\ref{e:momentum}), we require only the magnetic
induction equation to close the system. In the geometry considered here, the induction
equation equation~(\ref{e:induction}) is
\begin{equation}
  \ddt B =   -\mbfnabla\cdot(\mbfui B) +\lO\nabla^2 B,
  \label{e:inductionfull}
\end{equation}
where the second term on the RHS has the same form as Ohmic diffusion. We tune this
last term so that it dominates the numerical diffusion.

\subsection{Linear Theory\label{sss:dispersionrelation}}

Equations~(\ref{e:momentum}) and (\ref{e:inductionfull}) can be linearized to describe
small perturbations about a uniform equilibrium state. The analytical solutions 
provide physical insight and are useful for numerical tests. Assuming the 
perturbations are Fourier modes which depend on 
$(x,y,t)$ as $\exp{[i(k_xx+k_yy-\omega t)]}$ yields the dispersion relation
\begin{equation}
  \omega = -\f{\imath\nuin}{2} \pm \sqrt{k^2\,c_{Ai}^2 -(\f{\nuin}{2})^2},
  \label{e:disprel}
\end{equation}
where $k^2\equiv k_x^2+k_y^2$ and
$c_{Ai}=B/\sqrt{4\pi\rho_i}$ is the Alfv\'{e}n speed in the plasma.

If $k c_{Ai}\gg \nuin/2$, the roots of equation~(\ref{e:disprel}) describe forward
and backward propagating magnetosonic waves damped by ion-neutral friction
\begin{equation}\label{e:weakdamping}
\omega\approx\pm kc_{Ai}-\imath\f{\nuin}{2}.
\end{equation}
If $k c_{Ai} < \nuin/2$, $\omega$ is purely imaginary. In the limit
$k c_{Ai} \ll \nuin/2$, the roots of equation~(\ref{e:disprel}) are given
approximately by
\begin{eqnarray}
\omega&=&-\imath\nuin\label{e:strongdamping}\\
\omega&=&-\imath\f{k^2c_{Ai}^2}{2\nuin}\label{e:strongdamping2}.
\end{eqnarray}

Equations (\ref{e:weakdamping}) and (\ref{e:strongdamping}) can be compared to the
dispersion relation which accounts for ion feedback on
the neutrals \citep{KUP1969}. According to this more complete treatment, $\omega$ is
purely imaginary for $2\nuin\sqrt{\rho_i/\rho}<kc_{Ai}<\nuin/2$. The upper limit
is exactly what is predicted by equation~(\ref{e:disprel}), and
equation~(\ref{e:weakdamping}) agrees well with the exact dispersion
relation. The lower limit is not predicted by equation~(\ref{e:disprel}). At $k$ values
well below this lower cutoff, the wave propagates at the bulk Alfv\'en speed $B/
\sqrt{4\pi\rho}$. In this regime, the neutrals have time to be accelerated by the
ions within one wave period. This does not happen in our model because we omit
the drag force on the neutrals.

\begin{figure}[t]
  \plotone{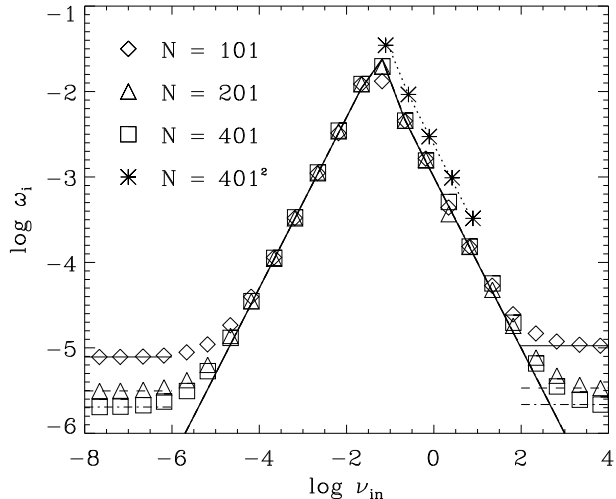}
  \caption{\label{f:disprel}Decay rate $\omega_i$ against collision
          frequency $\nuin$, according to equation~(\ref{e:disprel})
          for the 1D case (solid line) and  the 2D case (dashed line). 
          The symbols denote
          numerical results at resolutions as labeled in the plot. Note that
          the 2D case is shifted by exactly a factor of $2$ with respect to the 1D
          result. The flat parts at low and high $\nuin$ are caused by
          numerical diffusion, leaving roughly two orders of magnitude in $\nuin$
          in the declining branch of $\omega_i$ (eq.~[\ref{e:strongdamping2}])
          for numerical studies.}
\end{figure}

The solid line in Figure \ref{f:disprel} shows the predicted dependence of the damping rate 
(i.e. the imaginary part of the wave frequency $\omega$ of eq.~[\ref{e:disprel}]) 
on $\nuin$. The damping mode changes at $2c_{Ai} k$ as expected.
A discussion of the overplotted numerical results is left to \S\ref{ss:scheme}.

\subsection{The Neutral Flow\label{ss:gpflow}}

We adopt a
two-dimensional version of the divergence free, ``circularly polarized'' flow of
\citet{GAP1992}, (hereafter GP-flow), which we write in component form as
\begin{eqnarray}
  u_{n,x} = \sqrt{\f{3}{2}}\vGP
            \cos(2\pi(\kGP y+\epsilon\sin(2\pi\kGP \vGP t)))\nonumber\\
  u_{n,y} = \sqrt{\f{3}{2}}\vGP
            \sin(2\pi(\kGP x+\epsilon\cos(2\pi\kGP \vGP t))).
  \label{e:gpflow}
\end{eqnarray}
We have chosen this flow because it has stretching properties
(\citet{GAP1992} show it to be a dynamo flow) and can be written in
closed form. Its degree of chaos can be tuned through the choice of the parameter
$\epsilon$.\footnote{Note that $\epsilon$ contains a factor of $2\pi$ in our definition
(eq.~\ref{e:gpflow}), thus being by that factor larger than the
definition in \citet{GAP1992}.}
For $\epsilon=0$ the flow is steady and possesses cellular structure at a
single spatial scale. The corners of the cells are hyperbolic stagnation points near
which the fluid undergoes exponential expansion along one axis and exponential 
compression along the other. Fluid circulates steadily around the center of its cell 
with an eddy turnover time of order
\begin{equation}
  \tGP = \left(\kGP\vGP\right)^{-1}.
  \label{e:eddyturnover}
\end{equation}
Since each eddy retains its amount of fluid, we expect turbulent transport to be minimal.

The flow pattern for $\epsilon > 0$ can be visualized as eddies at a single scale
travelling in snake-like patterns across the domain. The position of each eddy 
oscillates by $\epsilon/\kGP$. For $\epsilon\ll 1$, this is also roughly the size 
of the chaotic region. Advected quantities are not bound to the eddies,
but travel from one to another, leading to turbulent transport. 

\begin{figure}[t]
  \plotone{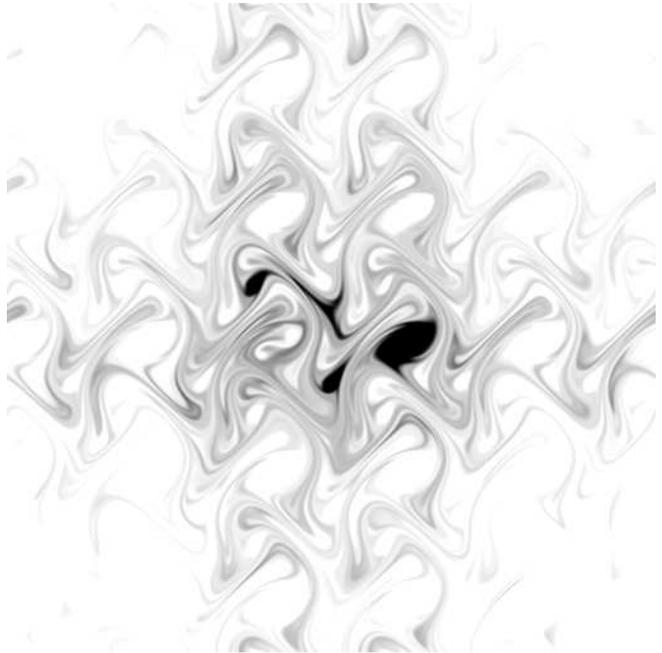}
  \caption{\label{f:prettypicture}Tracer pattern caused by GP-flow after
           $t=2\tGP$. At $t=0$, a tracer quantity is deployed
           into a circular region with diameter $L/4$.
           The GP cell number $\kapGP = 5$, and $\epsilon = 0.3$. 
           Resolution $N=1601^2$.}
\end{figure}

Figure~\ref{f:prettypicture} shows an example of the GP-flow for
$\kapGP$, the number of cells per domain length $L$, equal to 5 and $\epsilon = 0.3$. 
At $t=0$, we deploy a tracer quantity into a circular region with diameter $L/4$. 
The resulting tracer distribution is shown at $t=2\tGP$.
We discuss the choice of $\epsilon$ in \S\ref{ss:gpproperties}.

\subsection{Parameter Regimes\label{ss:parameters}}

There are four diffusivities: the numerical diffusivity $\lnum$,
the magnetic diffusivity $\lO$ defined in equation~(\ref{e:inductionfull}), the ambipolar
diffusivity $\lAD$ defined in equation~(\ref{e:lAD}), and the eddy diffusivity of the GP 
flow $\led$, which is ${\mathcal{O}}(\vGP/\kGP)$ (see eq.~[\ref{e:led}]). Normalizing by 
$\led$ and using equation~(\ref{e:lAD}), we require
\begin{equation}\label{e:lambdaratio}
\f{\lnum}{\led}\ll\f{\lO}{\led}\ll\f{\kGP\cAi}{\nuin}\f{\cAi}{\vGP}\ll 1.
\end{equation}
If equation~(\ref{e:lambdaratio}) is fulfilled then numerical diffusion is smaller than 
all other diffusion, Ohmic diffusion is weaker than ambipolar diffusion, as it is in the
interstellar medium, and eddy diffusion is faster than ambipolar diffusion, which allows us
to test the hypothesis that turbulence enhances the rate of magnetic field redistribution.
The last inequality is equivalent to requiring that the field is frozen to the eddies:
$\RAD\gg 1$ (see eq.~[\ref{e:RAD}]).

We also seek a separation of lengthscales in the problem. The largest scale possible in
a periodic domain is $L$, while the eddy scale is $L/\kapGP$. The distinction between large
and small scales can be maintained only if
\begin{equation}\label{e:kapGP}
\kapGP\gg 1.
\end{equation}

Finally, as we saw in equation~(\ref{e:disprel}), waves propagate only for sufficiently 
large wavenumbers. We are actually interested in the nonpropagation regime, because only 
in that case is the neutral flow imprinted on the ions. Thus, we require
\begin{equation}\label{e:nonprop}
\f{\kGP\cAi}{\nuin}\ll\f{1}{2}.
\end{equation}
When equation~(\ref{e:nonprop}) is combined with the last condition in equation
(\ref{e:lambdaratio}), the result is an upper limit on $\cAi/\vGP$. In the interstellar
medium, where the magnetic field is close to equipartition with the turbulence, this
quantity can be of order $(\rho/\rho_i)^{1/2}$, and hence quite large. For computational
reasons, we are unable to make $\kGP\vGP/\nuin$ as small as it should be, and therefore
$\cAi/\vGP$ will be smaller than it ought to be. However, there is no physical reason
why turbulence cannot obey the proper ordering.

\begin{figure}[t]
  \begin{center}
  \plotone{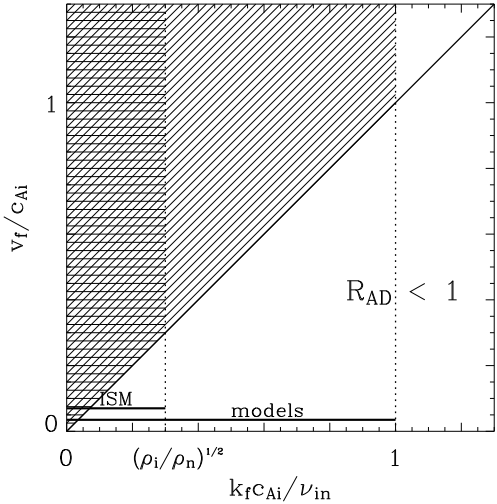}
  \end{center}
  \caption{\label{f:validregime}Valid regimes (schematically) for the
           ISM (horizontal lines) and the models (diagonal lines)
           under the strong coupling condition (eq.~[\ref{e:sca}]).
           Vertical lines denote the nonpropagation limit according to  
           equation~(\ref{e:nonprop}) for the ISM and the models.}
\end{figure}
The restrictions on $\vGP/\cAi$ and $\kGP\,\cAi/\nuin$ resulting from
equations~(\ref{e:lambdaratio}) and (\ref{e:nonprop}) are sketched
in Figure~\ref{f:validregime}.
For $\kGP\,\cAi/\nuin > 1$, we are in the weakly damped branch, and
anything below the diagonal thick line corresponds to $R_{AD} < 1$.

Taking these constraints together,
the range of permissible $\nuin$ is limited to at most 2 orders of magnitude, the range
of $B$ by slightly more than one order of magnitude, and $\kapGP$ between $5$ and $10$.

Tables~\ref{t:rad} and \ref{t:ladlo}, summarize the parameters for all models.
For each combination of $B$ and $\nuin$, we ran models for three GP forcing scales, namely
$\kapGP\in\{5,7,10\}$ at a resolution of $N=801^2$. These models were completed by a set of 
pure tracer models (i.e. models with $\nuin = 0$, and $B$ subjected to turbulent diffusion 
only). To assess resolution effects on the results, we repeated the calculations at 
$N=1601^2$. 
\begin{deluxetable}{ccccc}
  \tablewidth{0pt}
  \tablecaption{Ambipolar Reynolds number $R_{AD}$\label{t:rad}}
  \tablehead{\colhead{$\downarrow B\,\,\,\nuin\rightarrow$}&\colhead{$0.7$}&\colhead{$2.3$}&\colhead{$7.1$}&\colhead{$23.0$}}
  \startdata
  $0.1$ &$3.4\times10^3$&$1.1\times10^4$&$3.4\times10^4$&$1.1\times10^5$\\
  $0.5$ &$1.3\times10^2$&$4.5\times10^2$&$1.3\times10^3$&$4.5\times10^3$\\
  $1.0$ &$3.4\times10^1$&$1.1\times10^2$&$3.4\times10^2$&$1.1\times10^3$\\
  $2.0$ &$8.6\times10^0$&$2.8\times10^1$&$8.6\times10^1$&$2.8\times10^2$\\
  \enddata
\end{deluxetable}
\begin{deluxetable}{ccccc}
  \tablewidth{0pt}
  \tablecaption{Ratio $\lAD/\lO$\label{t:ladlo}}
  \tablehead{\colhead{$\downarrow B\,\,\,\nuin\rightarrow$}&\colhead{$0.7$}&\colhead{$2.3$}&\colhead{$7.1$}&\colhead{$23.0$}}
  \startdata
  $0.1$ &$4.8\times10^{-1}$&$1.4\times10^{-1}$&$4.8\times10^{-2}$&$1.4\times10^{-2}$\\
  $0.5$ &$1.2\times10^{1}$ &$3.6\times10^{0}$ &$1.2\times10^{0}$ &$4.0\times10^{-1}$\\
  $1.0$ &$4.8\times10^{1}$ &$1.4\times10^{1}$ &$4.8\times10^{0}$ &$1.4\times10^{0}$ \\
  $2.0$ &$1.9\times10^{2}$ &$5.8\times10^{1}$ &$1.9\times10^{1}$ &$5.8\times10^{0}$ \\
  \enddata
\end{deluxetable}

\subsection{Initial Conditions\label{ss:ic}}
We prepare our system with the initial conditions as given
in equation~(\ref{e:initcond}) for the 2D case.

\begin{eqnarray}
  B(x,y,0)  = B_0 + \f{1}{2}B_1\left[1+\cos(\kP x)\cos(\kP y)\right]\nonumber\\
  \rho(x,y,0)  = \rho_0 - \f{1}{2}\rho_1\left[1+\cos(\kP x)\cos(\kP y)\right].
  \label{e:initcond}
\end{eqnarray}
The 1D case is initialized similarly, with $y\equiv 0$. 
We add the offset in $B$ to prevent field reversals, as these are known to
lead to singularities in the presence of ambipolar drift \citep{BRZ1995,MAS1997}. We choose
the perturbation amplitudes to be $B_1 = 0.1B_0$, $\rho_1=0.1\rho_0$,
and the perturbation wave number $\kP$ to be $2\pi/L$, i.e. the perturbations 
reside on the largest possible mode. According to the formalism of QDT discussed in the 
Appendix, the $\kP$ components of $B$ and $\rho$ represent the large scale fields which are 
predicted to diffuse under the action of the small scale fields.

Ionization equilibrium (\S\ref{ss:assumptions}) implies that $\rho_i$ is slaved
to $\rho$. In the physical environments of interest here, $\rho_i\propto\rho^{1/2}$. Thus, 
the $10$\% perturbation of $\rho$ imposed here (eq.~[\ref{e:initcond}]) would cause a
$5$\% perturbation of $\rho_i$. For the sake of simplicity in the numerical scheme,
we kept $\rho_i$ constant in time and space. We believe the $5$\% variation in ion 
density to be too small to affect our results qualitatively.

To summarize, we initialize $B$ and $\rho$ using equations~(\ref{e:initcond}) and
evolve the system in space and time using equations~(\ref{e:momentum}) and 
(\ref{e:inductionfull}), where $\mbfun$ in equation~(\ref{e:momentum}) is given by
equation~(\ref{e:gpflow}).

%
%
\section{Numerics\label{s:numerics}}

\subsection{The scheme\label{ss:scheme}}
We solve equations~(\ref{e:momentum}) and (\ref{e:inductionfull}) using a modified
version of the 1st order gas-kinetic flux-splitting method described by
\citet{XUK1999} and \citet{TAX2000}.
MUSCL\footnote{Monotone Upwind Schemes for Scalar Conservation Laws} limiters
allow a 2nd order reconstruction of the flow variables at the cell walls.
As we are evolving only $\hat\mbfz B(x,y,t)$, $\mbfnabla\cdot\mbfB\equiv 0$
is satisfied. The (otherwise passive) $z$ velocity component
serves as a tracer in order to measure the transport properties of the flow.
The AD drag force terms are added as external source terms to the cell-centered
momenta and can be treated implicitly if necessary (similarly to
\citet{TOT1995,MAS1997}). The CFL-condition includes the diffusion timestep.
As we are interested in representing the peak value of the initial magnetic
field exactly, we employ a grid with an odd number of cells
(the symmetry axis of the initial magnetic field goes through a cell center).

Figure~\ref{f:disprel} serves as a validation of the method for our problem.
The decay of the magnetic field is measured in terms of the
peak magnetic field amplitude against time. The initial conditions for 1D
and 2D are given in equations~(\ref{e:initcond}).
Apart from the fact that the numerical data points agree with the predictions in
most cases at the 2\%-level, we note the flattening at low and high $\nuin$.
Both are caused by numerical diffusion. For small $\nuin$, this diffusion arises
from the limited grid resolution. At $\nu_{num}$, the point where the numerical
results start to leave the predicted curve, the physical diffusion becomes
smaller than the intrinsic numerical diffusion, which from then on governs the diffusive
behavior. As soon as the numerical curve flattens, the results are meaningless
for any physical problem. The timestep in this weakly damped branch is controlled
by the diffusion timestep as in e.g. \cite{MNK1995}. 

For large $\nuin$, the diffusion is caused by the finite timesteps. If the
collision time $\nuin^{-1}$ becomes smaller than the timestep $\Delta t$, the number
of collisions per $\Delta t$ is limited. In other words,
$\Delta t \ll \nuin^{-1}$ must be guaranteed in order to prevent losing
collisions and thus introducing numerical diffusion. The timesteps would be
impractically small, and pursuing the computations would require an implicit
treatment of the whole ion momentum equation. Although this is theoretically
possible, we chose to implement a 2nd order Runge-Kutta time stepping method which 
guarantees results unaffected by numerical diffusion up to $\nuin \approx 70$
(see Fig.~\ref{f:disprel}).

Note that the predictions for the 2D decay rate (dashed line) coincides exactly
with the numerical datapoints as well. It is this agreement which makes us
confident in using the method for the
investigation reported here. Further validation is presented in \S\ref{s:results}.

\subsection{Control of numerical diffusion\label{ss:controlnumdiff}}

As we mentioned in connection with equation~(\ref{e:inductionfull}), we
control the diffusive behavior of the scheme at grid scale by
introducing an artificial resistivity $\lO$ which dissipates the field above the grid scale.
As is usually the case in astrophysical computations, $\lO$ is much larger than the 
physical resistivity. However, it is necessary because the GP-flow mixes scales very 
efficiently, leading to diffusion at the smallest possible scale after approximately an
eddy-turnover time $\tGP$. This artificial resistivity
guarantees well-behaved diffusion properties above the grid
scale. It acts both on the magnetic field and the tracer field, which we identify
as $\rho$. Thus, $B$ and $\rho$ are guaranteed to be smoothed on the smallest scales
in the same manner.

\begin{figure}[t]
  \plotone{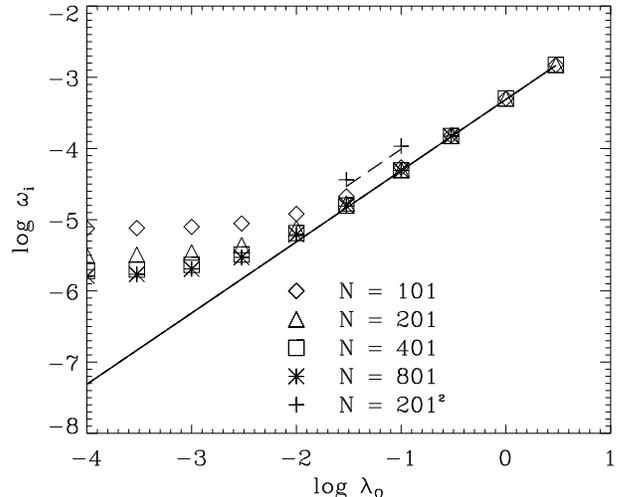}
  \caption{\label{f:resdisprel}Decay rate $\omega_i$ against Ohmic diffusivity
          $\lO$ according to equation~(\ref{e:resdisprel1d}) (solid line) for the
          1D case and (dashed line) for the 2D case. The symbols denote
          numerical results at resolutions as labeled in the plot. Note that the
          2D case is exactly a factor of $2$ shifted against the 1D result. The
          flat parts at low $\lO$ are caused by numerical diffusion.}
\end{figure}

In the flux-splitting scheme as described above, the amount of diffusion 
(last term of eq.~[\ref{e:inductionfull}]) is given
by the slope of $B$, which in turn is computed during the reconstruction of the
magnetic field at the cell walls. Thus we can implement the resistivity efficiently
in our scheme. Restricting ourselves for the moment to a purely
resistive medium -- excluding ambipolar diffusion -- the dispersion relation in the 
physically interesting regime (small $\lO$) is given in 1D by
\begin{equation}
  \omega = -\f{\imath\lO k^2}{2} \pm \sqrt{c_{Ai}^2 k^2-(\f{\lO k^2}{2})^2},
  \label{e:resdisprel1d}
\end{equation}
where we are interested in the case $\lO k^2 \ll c_{Ai}^2 k^2$. The corresponding
tests are shown in Figure~\ref{f:resdisprel}.
Numerical diffusion (flat part for small $\lO$) occurs at the same level
as for the implementation of ambipolar diffusion (see Fig.~\ref{f:disprel}).
Comparing the values of $\omega_i$ in Figure~\ref{f:resdisprel} with those of
Figure~\ref{f:disprel} shows that our parameter space is limited by
resistive diffusion.

%
%
\section{Results\label{s:results}}

We computed the evolution of $B$ and of a tracer field for a set of realizations
of the basic model described in \S\ref{s:description}; see Tables 1 and 2. We found that
some, but not all, aspects of the transport of these quantities can be cast in terms
of turbulent diffusion. After explaining how we measure the total diffusivity $D$ 
in our models (see Appendix), we discuss the transport properties of the GP-flow
in \S\ref{ss:gpproperties}. In \S\ref{ss:tracertransport} and 
\S\ref{ss:diffbaloney} we describe the evolution of the 
tracer $\rho$ and the transport of $B$, respectively. Turbulent transport affects
$\rho$ and $B$ in different ways, requiring a closer view on the ion flow
(\ref{ss:ionflow}). Finally in \S\ref{ss:diffbrho} we discuss the evolution of the flux 
to mass ratio $B/\rho$.

\subsection{Transport Properties of the GP-flow\label{ss:gpproperties}}

The transport properties of a flow are closely related to its Lyapunov exponents (see for
example \citet{DRA1992}). We used a method implemented by \citet{BCT2001}, based on
\citet{SOW1994} to calculate the Lyapunov exponents for the GP-flow. It
follows the time evolution of the separation $d$ of two points, initially
an infinitesimal distance $d_0$ apart. 
If one or both points are located in a chaotic region, the separation grows 
exponentially. The grid resolution (in this case $128^2$ and $256^2$) is given by the 
number of paired starting points for following the particle trajectories, in our case
distributed evenly over the flow domain.

\begin{figure}[t]
  \plotone{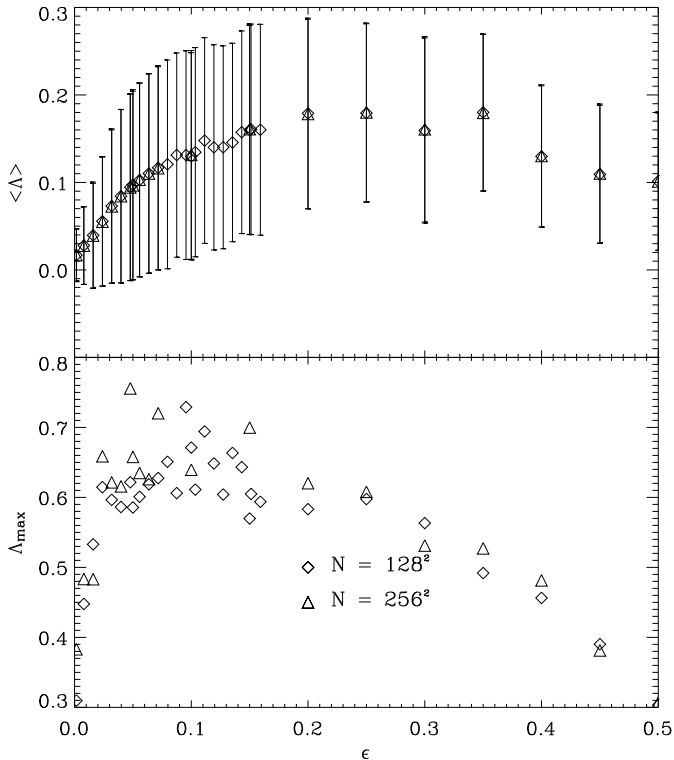}
  \caption{\label{f:lyapexp}Mean and maximum Lyapunov exponents against $\epsilon$.
           Error bars show the standard deviation across one frame. 
           The maximum exponent $\Lambda_{max}$ peaks at small $\epsilon$ because
           in this case, particle orbits are long and the chaotic regions are 
           located at the cell boundaries.
           The flow's transport properties are determined primarily by the mean exponent
           $\langle\Lambda\rangle$.} 
\end{figure}

Figure~\ref{f:lyapexp} gives the dependence of the domain averaged Lyapunov exponent
 $\langle\Lambda\rangle$ and maximum Lyapunov exponent $\Lambda_{max}$ on $\epsilon$.
The mean Lyapunov exponent $\langle\Lambda\rangle$ peaks near $\epsilon \approx 0.25$,
corresponding to a maximum phase shift of $\pi/2$ (see eq.~[\ref{e:gpflow}]). The maximum
exponent shows considerable scatter (mirrored in the large dispersions about the
mean), but it too drops nearly monontonically for $\epsilon \gtrsim 0.25$.  This came
somewhat as a surprise to us, as a rising $\epsilon$ increases the
time-dependence of the GP-flow.

\begin{figure}[t]
  \plotone{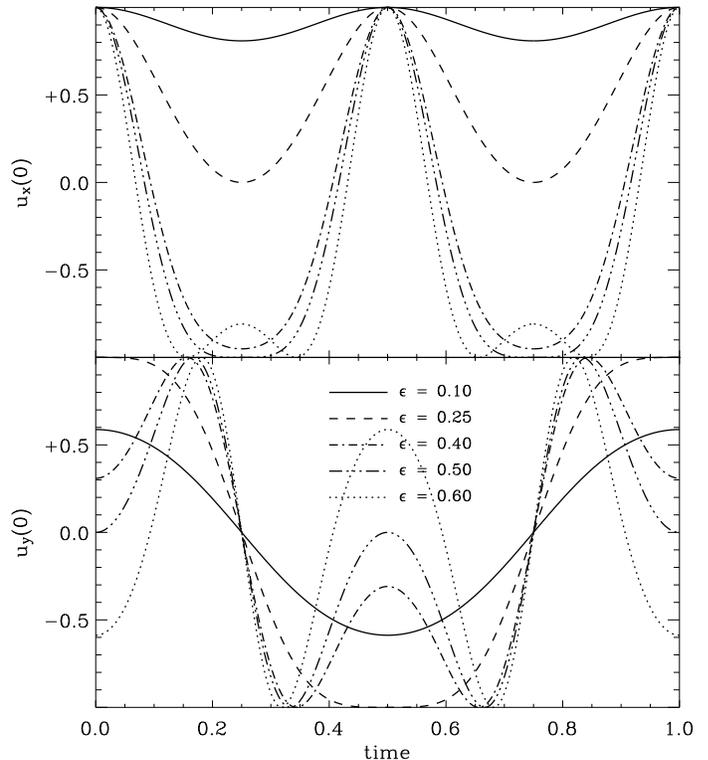}
  \caption{\label{f:gpexample}Velocities $u_x$ and $u_y$ according to
           equation~(\ref{e:gpflow}) at $x = y = 0$. For $\epsilon > 0.25$
           (maximum phase shift $\pi/2$), oscillations in $u_y$ change to the
           next higher octave, the same happens for $u_x$ for $\epsilon > 0.5$.
           By this, the coherent flow structure is disrupted.}
\end{figure}

Figure~\ref{f:gpexample} explains this behavior.
For $\epsilon > 0.25$ (maximum phase shift $\pi/2$), the frequency in velocity
perturbations is doubled, disrupting the coherent flow structures and reducing the distance
a test particle can be transported. Based on
this qualitative argument, we expect a maximum in the transport rate for
$\epsilon \simeq 0.25$.

\begin{figure}
  \plotone{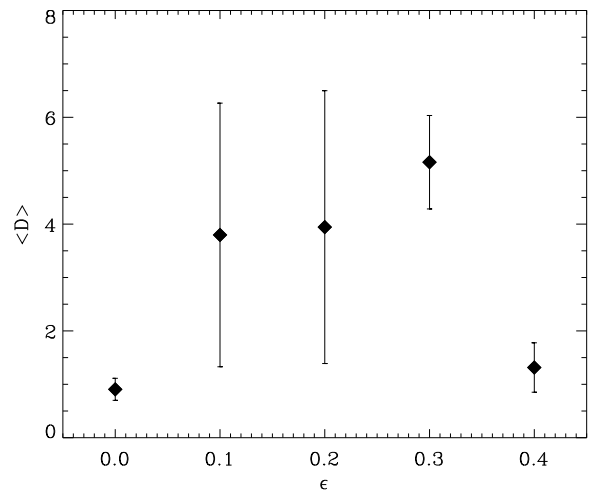}
  \caption{\label{f:diffepsilon}Diffusion constant $D$ against complexity
          parameter $\epsilon$ (eq.~[\ref{e:gpflow}]). The error bars give the
          standard deviations about the mean in equation~(\ref{e:compeffdiffusion}).}
\end{figure}

Figure~\ref{f:diffepsilon} demonstrates the connection between the complexity
parameter $\epsilon$ in the GP-flow (eq.~[\ref{e:gpflow}]) and the diffusion
constant $D$ (eq.~[\ref{e:compeffdiffusion}]). The diffusion constant peaks
approximately at the $\epsilon$ for which the mean Lyapunov exponent
$\langle \Lambda\rangle$ is maximal (see Fig.~\ref{f:lyapexp}). The
error bars reflect temporal fluctuations in $D$.
Because we wish to maximize the transport, we chose $\epsilon=0.3$ in all the models
discussed subsequently.

We  assessed the effect of the GP-flow's temporal periodicity on the transport rate by
running some numerical experiments in which we replaced the factors of $\kGP\vGP t$ in
equations~(\ref{e:gpflow}) by random phases chosen from a uniform distribution in ($0,2\pi$)
at fixed intervals in time. The resulting transport was less than for the GP flow with the
same value of $\epsilon$, so we pursued this model no further.

\begin{figure*}
  \begin{center}
  \includegraphics[width=15.0cm]{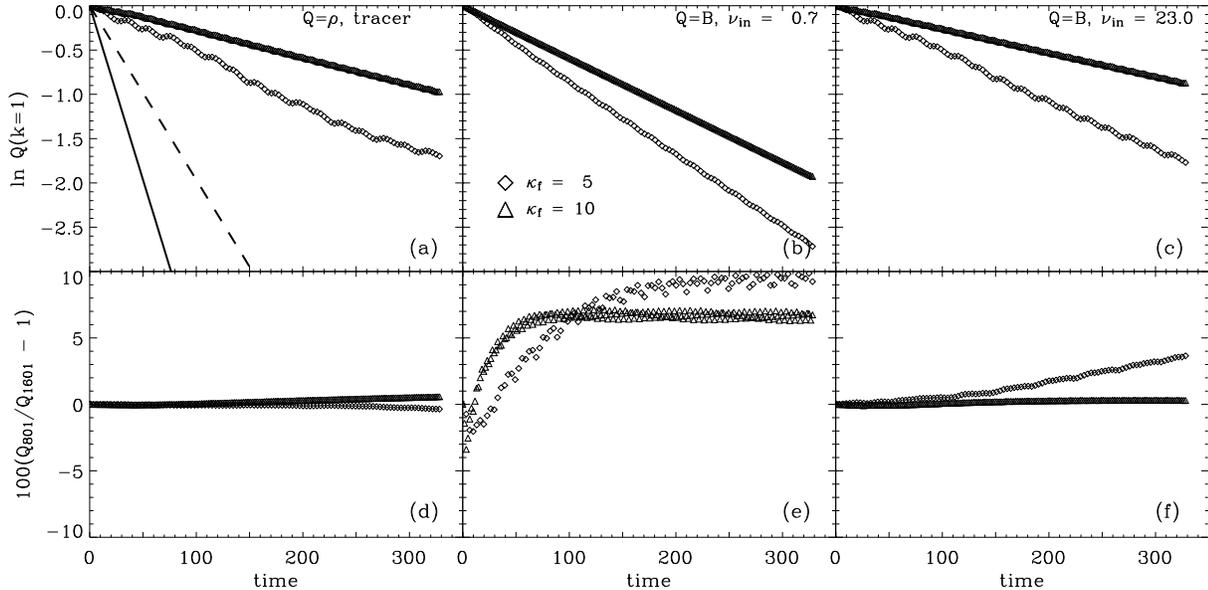}
  \end{center}
  \vspace{0.2cm}
  \caption{\label{f:decayall}$(k=1)$-mode of $Q$ -- where $Q$ is either
          the density $\rho$ or the magnetic field strength $B$ --
          against time for resolution $N=801^2$, $\kapGP\in \{5,10\}$ and 
          $\epsilon = 0.3$.
          (a) Decay of density field. The solid ($\kapGP=5$) and dashed ($\kapGP=10$) 
          lines stand for the turbulent decay rate expected from QDT (eq.~[\ref{e:led}]). 
          (b) Decay of magnetic field with $\nuin = 0.7$.
          (c) Decay of magnetic field with $\nuin = 23.0$.
          The slope gives the decay rate $\omega_i$.
          Panels (d) to (f) show the difference $(Q_{801}-Q_{1601})/Q_{1601}$ in percent
          for runs of resolutions $N=801^2$ and $N=1601^2$ and parameters corresponding
          to panels (a) through (c).
          For $\nuin = 0.7$, $B$ develops strong peaks, which tend to be underresolved
          at $N=801^2$.}
\end{figure*}

\subsection{Transport of a Tracer\label{ss:tracertransport}}
The transport of a passive scalar by the GP flow establishes a baseline against which the
transport of the magnetic field can be compared. It is also the relevant transport
equation for the gas density $\rho$. We compute the transport by integrating
equation~(\ref{e:inductionfull}) with $\mbfui\equiv\mbfuGP$. The
initial conditions are given by equations~(\ref{e:initcond}). Numerical diffusion is 
controlled as discussed in \S\ref{ss:controlnumdiff}

Panel (a) of Figure~\ref{f:decayall} shows the time evolution of the average of the Fourier
coefficients of $\rho_{k=1}$ (as defined following eq.~[\ref{e:mdiffint}])
for two GP cell sizes ($\kapGP=5,10$). 
The curves clearly demonstrate exponential decay. The
difference in decay rates between the $\kapGP=5$ and $\kapGP=10$ cases agrees reasonably
well with the expected $u^2\tau$ scaling of the turbulent diffusivity, since
doubling $\kapGP$ decreases $\tau$ by a factor of $2$ (eq.~[\ref{e:eddyturnover}]).
We note that the decay rates do not reach the values predicted by QDT 
(eq.~[\ref{e:led}])\footnote{This does not come as a surprise since QDT provides an 
estimate whose general assumptions are quite different from the assumptions made here: 
e.g, the GP-flow has a single scale and it is spatially and temporally periodic.
The oscillation of the eddies in position prevents the maximum stretching of fluid 
elements in GP-cell corners.}.
The bumps in the curves are caused by time variations in the GP flow. Not only does the 
temporal frequency of the bumps correspond to the GP frequency, but the difference between 
two spatial resolutions (Fig.~\ref{f:decayall}(d)) is very small over the duration of 
the calculation.

\begin{figure}
  \plotone{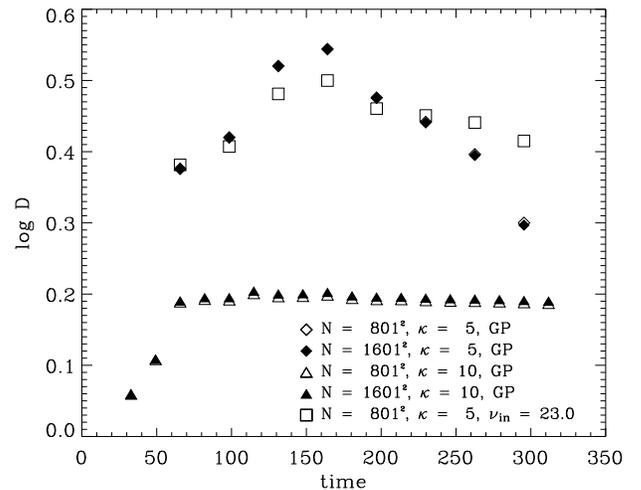}
  \caption{\label{f:diffconsttracer}Diffusion constant
           (eq.~[\ref{e:compeffdiffusion}]) for the turbulent decay of the
           passive scalar, identified with the density $\rho$.
           The open squares denote results for turbulent ambipolar diffusion
           at the highest collision frequency ($\nuin = 23.0$),
           demonstrating that for large collision frequencies
           the ions follow the GP-flow closely.}
\end{figure}

Figure~\ref{f:diffconsttracer} gives $D$ as defined in
equation~(\ref{e:compeffdiffusion}) for
the same GP tracer models as in Figure~\ref{f:decayall}. The settling effect
of larger GP cell numbers is clearly visible. Note again that the different
resolutions give identical results.

\begin{figure*}
  \begin{center}
  \includegraphics[width=15.0cm]{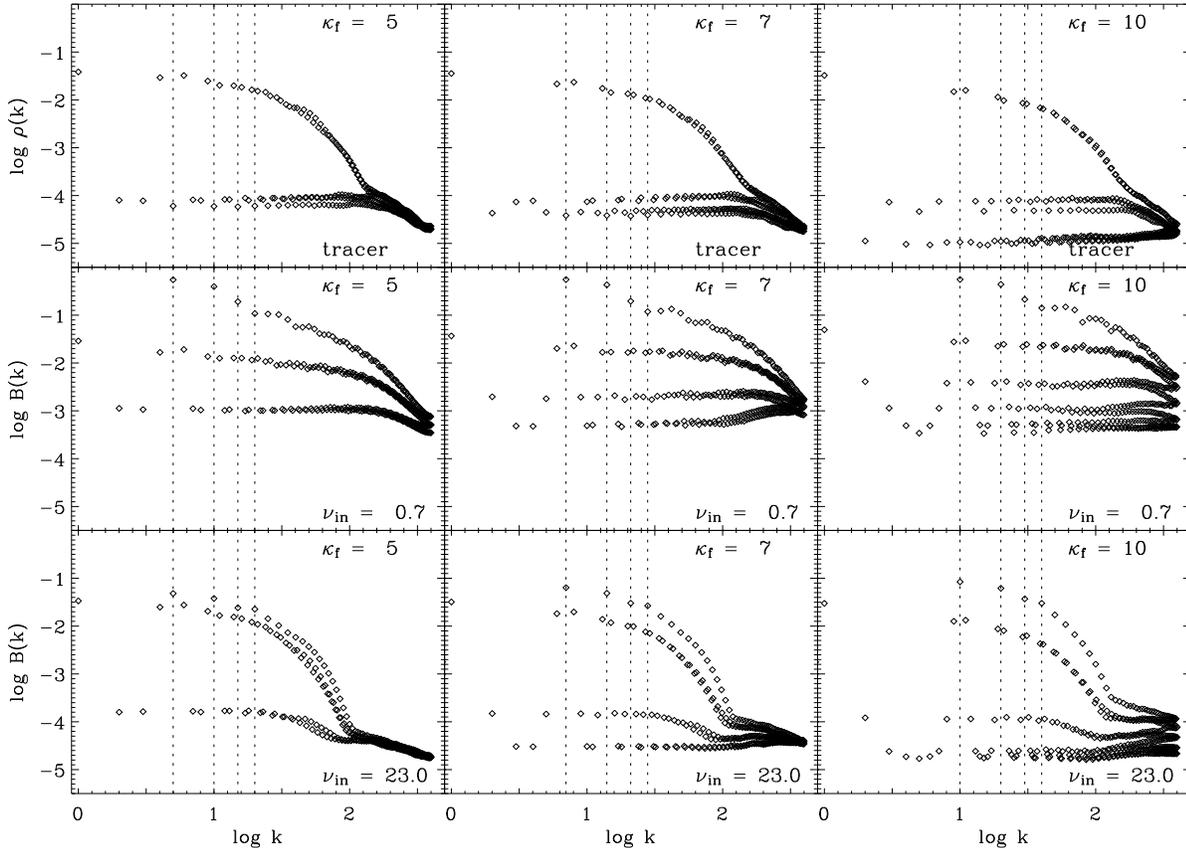}
  \vspace{0.2cm}
  \end{center}
  \caption{\label{f:magfourier}Fourier spectra of $\rho$ (upper row) and $B$ (lower rows
          for minimum and maximum collision frequency $\nuin$)
          for $\kapGP\in\{5,7,10\}$. Vertical dashed lines denote multiples of $\kapGP$
          (see text). Resolution $N=801^2$.
          Each of the nine plots contains one spectrum. The Nyquist $k_N = 400$.}
\end{figure*}

The top row of panels in Figure~\ref{f:magfourier} show Fourier spectra of $\rho$ for
$\kapGP\in\{5,7,10\}$. The vertical dashed lines indicate multiples of $\kapGP$. 
All spectra display at least two branches with significantly different power. The
values in the maximum branch group at $n\kapGP\pm 1$, with $n\in\{1,2,...\}$.
This is a direct consequence of 
the nature of the advection operator. The velocity $\mbfu$ with its single scale $\kapGP$
initially beats together with the density perturbation, which is at $k=1$ (see
eq.~[\ref{e:initcond}]). This creates density structure at $\kapGP\pm 1$. Advection of
this secondary structure generates power at $2\kapGP\pm 1$, and so on. The spectra should
have power at these wavenumbers only; the power seen in Figure~\ref{f:magfourier} at other
$k$ (all branches except the maximum branch) is entirely due to numerical 
noise\footnote{Although eqs.~(\ref{e:momentum}) and
(\ref{e:inductionfull}) are solved in double precision, the Fourier transform is
computed in single precision.}. 
The amplitude of the noise is 2-3 orders of magnitude less than the power at wavenumbers
$n\kapGP\pm 1$ (the maximum branch) up to the Nyquist $k$, which we regard as acceptable. 

\subsection{Transport of the magnetic field\label{ss:diffbaloney}}

Panels (b) and (c) of Figure~\ref{f:decayall} show the decay of the
($\pm 1,\pm 1$) Fourier coefficients of $B$ for two resolutions and two collision 
frequencies. The decay rates, when averaged over the
GP flow period, are well fit by exponentials, and the scaling of the rates with $\kapGP$ is
consistent with turbulent diffusion. The decay rate at the higher collision frequency is
close to the GP rate shown in the leftmost panel. At the lower collision frequency, the
field decays substantially faster because of the relatively large value of $\lAD$ in 
that case: the turbulent and laminar diffusivities are additive.
The difference between these cases reflects the greater degree to
which the ion flow is slaved to the neutral flow in the high collision frequency case.

Panels (e) and (f) of Figure~\ref{f:decayall} give the differences between these
calculations for two resolutions. At the higher collision frequency, the differences
increase gradually with time and remain small over the duration of the calculation (although
they are larger than the corresponding differences for the tracer $\rho$ shown in the lower
left panel). At the low collision frequency, the difference grows quickly and saturates at
about 10\%. This shows the increased role of small scale structure in $\mbfui$ at low 
collision frequencies. Smaller $\nuin$ lead to larger compression and thus stronger peaks
in $B$, which in turn tend to be underresolved at $N=801^2$.

\begin{figure}[t]
  \plotone{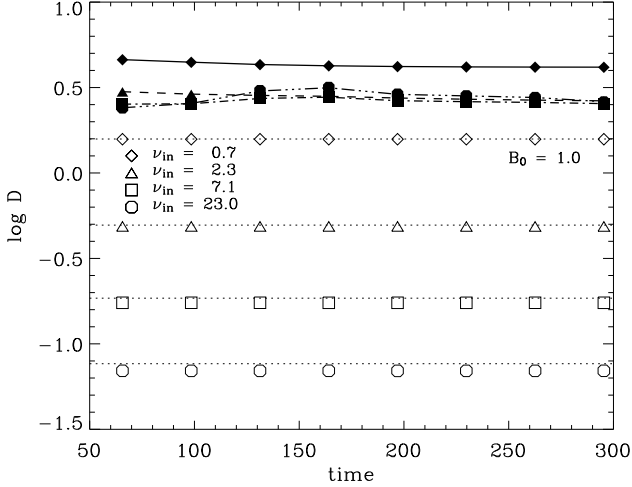}
  \caption{\label{f:diffconstadb10}Diffusivities $D$ against time according to
           equation~(\ref{e:compeffdiffusion}), for quiescent decay (open symbols)
           and turbulent decay (filled symbols for $N=801^2$, thick lines for
           $N=1601^2$). The quiescent diffusivities correspond to the theoretical 
           predictions (dotted lines), and the turbulent diffusivities are nearly 
           constant for the larger $\nuin$.}
\end{figure}

The diffusivities $D$ computed according to equation~(\ref{e:compeffdiffusion}) for the two
resolutions are nearly indistinguishable (Fig.~\ref{f:diffconstadb10}). The good agreement 
between the theoretically predicted quiescent diffusion rates $\lAD$ and the numerical 
result validates our method of measuring $D$ as well as provides an additional test of 
the underlying numerics. The approximate invariance of $D$ over time, and  
insensitivity of $D$ to $\nuin$, show that the transport of $B$ is well described as 
turbulent diffusion (the small offset between the turbulent diffusivity for $\nuin=0.7$ and 
the other cases is again caused by the relatively large value of $\lAD$ in that case). 
\begin{figure}[t]
  \plotone{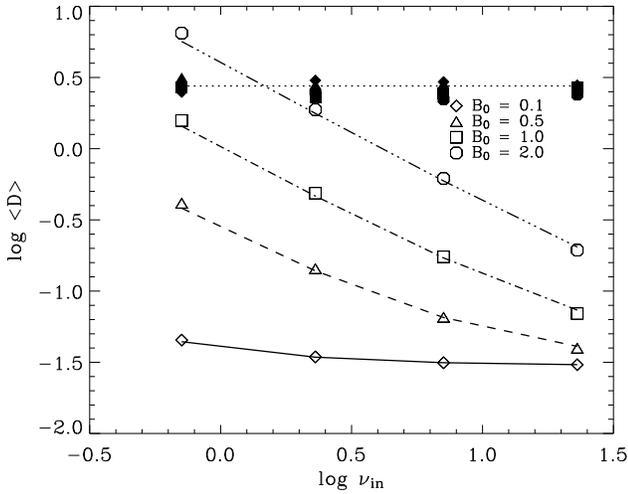}
  \caption{\label{f:diffconstadall}Diffusivities $D$ as in
           Figure~\ref{f:diffconstadb10}, but averaged over time, and for
           various field strengths. Open symbols give again quiescent rates
           and filled symbols stand for turbulent rates with the quiescent rates
           subtracted (see text). Note that the modelled
           quiescent rates reproduce the analytical predictions from
           $\lAD + \lO$ (thick lines). 
           The horizontal dotted line gives the turbulent diffusion rate
           for $\rho$, i.e. the pure GP-transport diffusion.}
\end{figure}

Because the ion flow is modified by Lorentz forces, we investigated the dependence of $D$
on magnetic fieldstrength. Figure~\ref{f:diffconstadall} plots the time-averaged $D$
-- denoted $\langle D\rangle$ -- for four collision strengths and four magnetic field 
strengths. The laminar diffusivities (lines) follow the expected dependence of diffusivity 
on $B$. In contrast, the turbulent diffusivities (calculated by measuring $D$ in a 
turbulent model and subtracting $\lAD$) depend only marginally on $B$, despite the fact 
that the Alfv\'{e}n Mach number varies from less than a tenth to slightly greater than 
unity. Note that at the lowest value of $B$, the quiescent diffusion rates are nearly 
independent of $\nuin$: at this value of $B$, the quiescent 
diffusivities are completely controlled by $\lO$. On the other hand, at the highest value 
of $B$ and lowest value of $\nuin$, the laminar diffusivity dominates the turbulent one. 
This case is less interesting to us for studies of the ISM.

The Fourier spectra of $B$ are shown in the second and third rows of 
Figure~\ref{f:magfourier} for two collision frequencies and $\kapGP\in\{5,7,10\}$. In 
striking contrast to the spectra of $\rho$ (top row of this figure), the spectra peak at 
multiples of $\kapGP$. This effect is particularly strong at the low collision frequency, 
and provides graphical evidence that the transport of $B$ to small scales is quite 
different from that of $\rho$. The disparity must be caused by differences between $\mbfui$ 
and the GP flow, i.e. the ion-neutral drift. To see this, we rewrite the induction 
equation~(\ref{e:inductionfull}), with resistivity included, in the form
\begin{equation}\label{e:induction2D}
\left(\f{\partial}{\partial t} + \mbfui\cdot\mbfnabla - \lO\nabla^2\right)B
=-B\mbfnabla\cdot\mbfui.
\end{equation}
The second term on the LHS of equation~(\ref{e:induction2D}) represents advection. It is 
predominantly, although not entirely, advection by the GP flow. As such, it cascades power 
to $n\kapGP\pm 1$, just as occurs for the tracer. The RHS of equation~(\ref{e:induction2D})
represents compression. Here the GP flow plays no role. If the inhomogeneous part of $B$ were
less than the mean part, the RHS could be approximated by $B_0\mbfnabla\cdot\mbfui$ and the 
Fourier spectrum of $\mbfnabla\cdot\mbfui$ would map directly to that of $B$. Since $B$ is 
roughly twice $B_0$, this is only part of the story; the compression term gives significant 
nonlinear coupling. In any case, it is clear that we must investigate the ion flow.

\subsection{The Ion Flow\label{ss:ionflow}}

The ion flow is driven by friction with the neutrals, by magnetic pressure, and by its 
own Reynolds stress. Friction drives $\mbfui$ toward $\mbfuGP$.  The Reynolds stress is 
nonlinear, and drives compressive flow at multiples of $\kapGP$. The magnetic field begins 
with power in the $(1,1)$ and $(0,0)$ components. This power is cascaded to smaller 
scales by advection, but is also coupled to the compressive part of the ion flow. 
The Lorentz force itself is nonlinear, creating additional harmonics in the ion flow.

\begin{figure}[t]
  \plotone{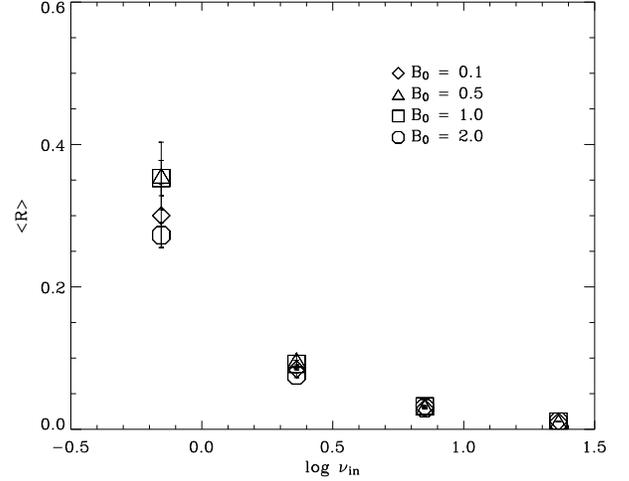}
  \caption{\label{f:ratdivcurl}Mean ratio of rms divergence over rms curl of $\mbfui$,
           $R=|\mbfnabla\cdot\mbfui|/|\mbfnabla\times\mbfui|$ against $\nuin$ for background
           field strengths $B_0$ as denoted in the plot. Error bars denote the standard
           deviation about the mean. Means were taken over the full computational domain.}
\end{figure}

We measure the compressibility of the flow through the parameter $\langle R\rangle$,
the ratio of the rms divergence of $\mbfui$ to its rms curl. Figure~\ref{f:ratdivcurl} shows
$\langle R\rangle$ as a function of $\nuin$ for four magnetic fieldstrengths.
\begin{figure*}
  \begin{center}
  \includegraphics[width=15.0cm]{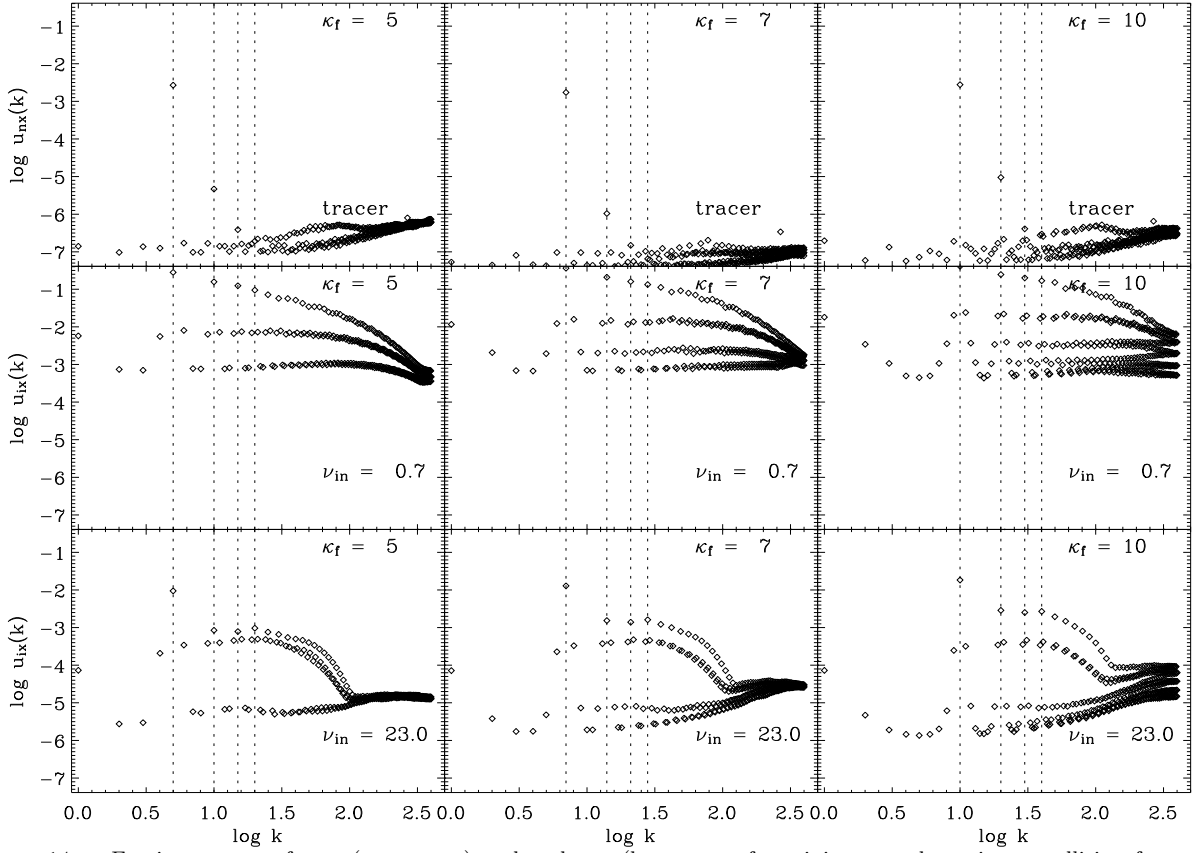}
  \vspace{0.2cm}
  \end{center}
  \caption{\label{f:vxfourier}Fourier spectra of $u_{nx}$ (upper row) and
          and $u_{ix}$ (lower rows for minimum and maximum collision frequency 
          $\nuin$) for $\kapGP\in\{5,7,10\}$. Vertical dashed lines denote multiples of 
          $\kapGP$ (see text). Resolution $N=801^2$.}
\end{figure*}
\begin{figure*}
  \begin{center}
  \includegraphics[width=15.0cm]{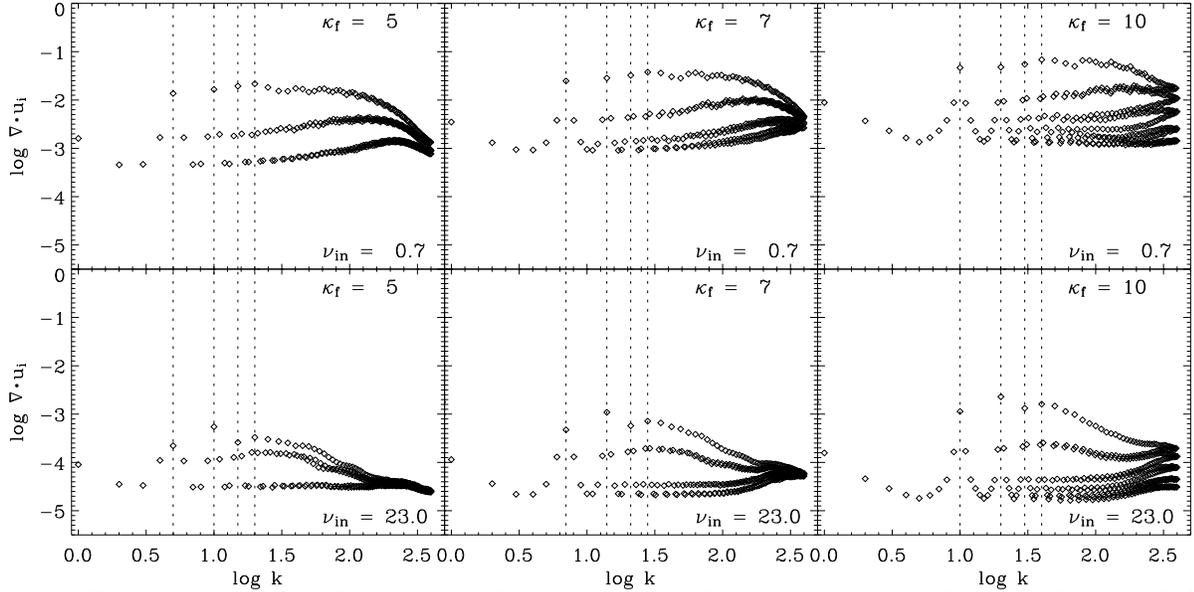}
  \vspace{0.2cm}
  \end{center}
  \caption{\label{f:divvfourier}Fourier spectra of $\mbfnabla\cdot\mbfui$
          for minimum and maximum collision frequency $\nuin$ and
          $\kapGP\in\{5,7,10\}$. Vertical dashed lines denote multiples of $\kapGP$.
          Resolution $N=801^2$. By definition, $\mbfnabla\cdot\mbfun\equiv 0$.}
\end{figure*}
At the lowest collision frequency the divergence is a few tenths the curl,
but quickly decreases with increasing $\nuin$. The value of $B$ influences
$\langle R\rangle$ less than the value of $\nuin$ does, and $\langle R\rangle$ is not 
monotonic in $B$. At low $B$, Lorentz forces are insignificant and compression arises 
from the Reynolds stress. As $B$ increases, Lorentz forces begin to play a role and lead 
to some additional compression. As $B$ increases further, the field resists compression and 
actually reduces $\mbfnabla\cdot\mbfui$.

The spectra of the $u_{ix}$ and $\mbfnabla\cdot\mbfui$ are shown in 
Figures~\ref{f:vxfourier} and \ref{f:divvfourier} . There is substantial power in both 
quantities at $\kapGP$ and its multiples. This can be attributed primarily to
frictional driving by $\mbfuGP$ with generation of power at the multiples of $\kapGP$ by 
$\mbfui\cdot\mbfnabla\mbfui$. The peak in the spectrum of  $u_{ix}$ at $\kapGP$ reflects 
the close correspondence between the ion flow and the GP flow. The spectrum of  
$\mbfnabla\cdot\mbfui$, however, is peaked not at $\kapGP$, but at its second or fourth 
multiple. The lower the collisionality, the higher the peak. The flow generated by the 
nonlinear Reynolds stress as the GP flow beats together with itself is compressive, and 
these flows themselves generate higher harmonics yet. If there were no Lorentz
force, the spectrum of $\mbfui$ would consist purely of harmonics of the GP flow. 
The magnetic field, however, is cascaded by nonlinear advection to the sideband 
wavenumbers $n\kapGP\pm 1$. The resulting power in the magnetic pressure gradient fills 
in the spectrum of $\mbfui$ at other wavenumbers.

In the interstellar medium, the respective magnitudes of the Reynolds stress and Lorentz 
force terms may well be reversed. As we commented below equation~(\ref{e:nonprop}), 
numerical considerations compel us to make $\vGP/c_{Ai}$ unrealistically large. However, 
both these terms are nonlinear, and both cascade the ion flow to small scales.


\subsection{Diffusion of $B/\rho$\label{ss:diffbrho}}

\begin{figure}[t]
  \plotone{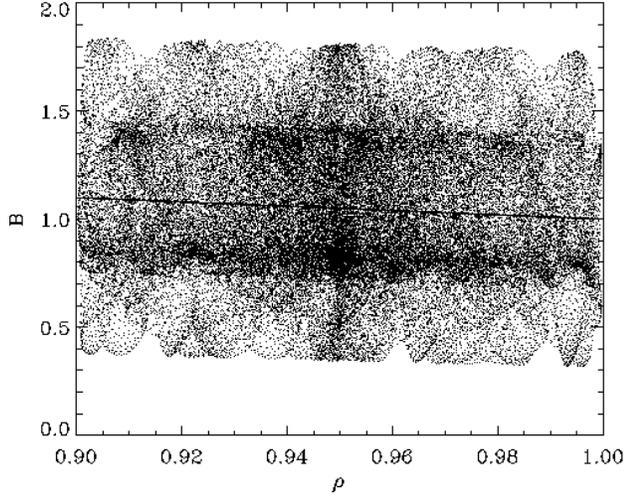}
  \caption{\label{f:scatterbrho}Magnetic field strength $B$ against
          density $\rho$ after one eddy turnover time $\tGP$ for
          $\kGP=5$ and $\nuin=0.7$. The line denotes the initial condition,
          and simultaneously the result when 
          $B$ is subjected to turbulent diffusion only, but not AD, after
          one $\tGP$.
          }
\end{figure}

Finally, we turn to the $B-\rho$ relation itself. A direct demonstration of how quickly 
$B/\rho$ changes for the GP-flow in combination with AD is given in 
Figure~\ref{f:scatterbrho}.
After one eddy turnover time, the initially strict correlation between
$B$ and $\rho$ (solid line) is completely destroyed. The solid line also represents
the correlation $B(\rho)$ for $B$ subject to the GP-flow alone, not including AD. 
In that case, $B$ and $\rho$ are both tracer fields, and diffuse in the same manner.

\begin{figure}[t]
  \plotone{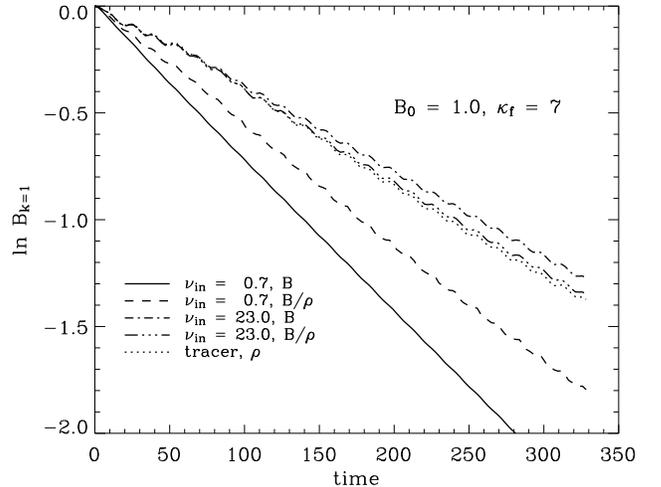}
  \caption{\label{f:decaybrho}Decay of the $(k=1)$-component of $B$ and $B/\rho$ against time
          for $\nuin\in\{0.7,2.3\}$.}
\end{figure}

The separation of $B$ and $\rho$ on small scales is reflected in the decay of the 
largescale, or ($\pm 1,\pm 1$) component of $B/\rho$ (Fig.~\ref{f:decaybrho}).
Since -- neglecting numerical diffusion -- $\rho$ ranges only from $0.9$ to $1.1$, 
we have to a fairly good approximation
\begin{equation}\label{e:Brhoapprox}
\left(\f{B}{\rho}\right)_{k=1}\approx\f{B_{k=1}}{\rho_0}
-\f{B_0}{\rho_0}\f{\rho_{k=1}}{\rho_0}.
\end{equation}
Since both $B_{k=1}$ and $\rho_{k=1}$ decay (see Fig.~\ref{f:decaybrho}),
it is clear that $(B/\rho)_{k=1}$ must decay as well, at least in the limit represented by 
equation~(\ref{e:Brhoapprox}). The decay rate estimated 
by equation~(\ref{e:Brhoapprox}) approximates the actual decay rate quite well.

%
%
\section{Summary\label{s:summary}}

The magnetic fieldstrength - density relation in the interstellar medium is observed to be 
quite weak, particularly in the atomic component and through the transition from diffuse 
gas to GMCs. The flatness of the trend is consistent with field-aligned flow or with an 
enhanced diffusion rate. This paper is concerned with testing the hypothesis that the 
diffusion rate is due to ambipolar drift accelerated by turbulence. We report on a 
series of numerical experiments intended to complement, and in some ways extend, 
analytical work on this problem by \citet{FAD2002}, \citet{KID2002}, and \citet{ZWE2002}.
Within the framework of interstellar medium physics, the setup is best imagined as small
scale turbulence acting on a relatively well ordered field.

We assumed a 2.5D geometry in which the motions are perpendicular to the magnetic field
$\hat\mbfz B$ and independent of $z$. In this situation, the fieldlines are shuffled but 
not bent, and there is no magnetic tension force. This type of turbulence is expected in 
a strong magnetic field \citep{STR1976} or as the outcome of an MHD cascade
\citep{GOS1995}, although in completely suppressing variation with $z$ we have taken
this description to an extreme degree.

We imposed a flow of the type considered by \citet{GAP1992} (see eq. [{\ref{e:gpflow}]) on
the neutrals. The GP flow is  spatially and temporally periodic, divergence free, and 
chaotic. This flow lacks some obvious features of interstellar turbulence, notably 
compressibility and a full spectrum of spatial scales. We chose this flow because it can 
be written in closed form, its level of chaos can be tuned by specification of a single 
parameter, and it is known to produce eddy diffusion. This made it a good candidate for 
a first series of numerical experiments. The basic assumption that the neutral flow can be 
prescribed as independent of Lorentz forces can be justified in either one of two ways. If 
the magnetic field were well below equipartition with the turbulent flow, this assumption 
would be valid independent of scale. This is, however, generally not the situation in the 
interstellar medium. Alternatively, we could restrict the computation to eddy sizes less 
than the cutoff wavenumber for strongly coupled MHD wave propagation; 
$k > 2(\rho_i/\rho_n)^{1/2}\nuin/c_{Ai}$
We emphasize, however, that this restriction is only a self consistency requirement for 
our calculation, not for turbulent ambipolar diffusion itself. 

We assumed a weakly ionized gas  in which the bulk density $\rho$ is advected with
the flow equal to the GP flow and the ion density $\rho_i$ is kept constant 
(see \S\ref{ss:ic}). We solved the ion momentum equation (eq.~[\ref{e:momentum}]) 
including the full nonlinear advection operator, the magnetic pressure gradient, and 
friction with the neutrals. This allowed us to go beyond the strong coupling 
approximation (eq.~[\ref{e:sca}]) made in the analytical work. However, we chose the GP 
wavenumber $\kapGP$ low enough that there would still be substantial friction in an eddy 
turnover time (eq.~[\ref{e:nonprop}]). The magnetic field is advected and compressed by 
the ion flow, and subject to a small amount of resistive diffusion, according to 
equation~(\ref{e:inductionfull}).

Our primary consideration, in choosing the computational scheme, was to minimize the role of
numerical diffusion.  Tests of the code based on linear theory, on  measured rates of laminar
diffusion, and with regards to resolution are shown in Figures~\ref{f:disprel}, 
\ref{f:decayall} and \ref{f:diffconstadb10}. By scrupulous choice of parameter regime we 
are able to keep the anomalous growth of small scale structure to a few percent or less 
while still considering collision frequencies varying by a factor of slightly more than 
$30$, magnetic fieldstrengths over a factor of 20, and up to a factor of 10 between the 
large scale quantities to be diffused and the GP eddies.

The  most serious physical shortcomings of the model, as a realization of interstellar 
turbulence, are its restricted geometry, kinematic prescription for the neutral flow, and 
relatively large ratio of turbulent speed to ion Alfv\'en speed. Although all three of 
these features were to some extent forced upon us by numerical considerations, we see some 
advantages to the first two, which have enabled us to study turbulent ambipolar diffusion 
in a relatively uncomplicated setting without a bevy of competing physical effects. The 
main consequence of the third feature is that it makes the Reynolds stress overly prominent 
in comparison to the Lorentz stress. However, even when $\mbfnabla\cdot\mbfui$ is 
relatively small, the differences between $\mbfui$ and $\mbfuGP$ are significant enough on 
small scales to decorrelate $B$ and $\rho$ (Fig.~\ref{f:ratdivcurl}).

These are the main results of our calculations:
\begin{enumerate}
  \item The GP flow causes decay of the large scale component of the density (or any other
	tracer field). The decay is brought about by mixing to small scales. Despite the 
        relatively small number of eddies in the system, the decay rate is well described 
        by a turbulent diffusivity of order $\vGP/\kGP$, as expected from mixing length 
        theory. The decay of the tracer is shown in Figure~\ref{f:decayall}.

  \item The ion flow driven by the GP flow causes decay of the largescale component of $B$.
        At the largest value of $\nuin$ in our experiment, the decay rate is close to the 
        decay rate of the tracer. At the smallest value, it is faster. We attribute this,
        at least in part, to the substantial role of laminar diffusion in this case.
        The eddy diffusivity is nearly independent of the 
        microscopic diffusivities (Ohmic and ambipolar), and of the numerical resolution.
        The decay of the field is shown in Figure~\ref{f:decayall}.

  \item Although the large scale structure in $\rho$ and $B$ decay at similar rates, the
        large scale structure in their ratio, $B/\rho$, decays at the eddy rate as well, 
        as shown in Figure~\ref{f:decaybrho}. This is brought about by the differences 
        between $\mbfuGP$ and $\mbfui$ on small scales. These small scale relative drifts 
        rapidly destroy any correlation between $B$ and $\rho$, as shown in 
        Figure~\ref{f:scatterbrho}. Thus, neither point to point measurements nor line
        of sight averages would yield a $B - \rho$ relation. We regard the separate 
        transport of $B$ and $\rho$ as Eulerian transport, and the decay of $B/\rho$ as 
        Lagrangian transport.
\end{enumerate}

What are the implications for the interstellar medium? The enhanced diffusion rate 
demonstrated here must be balanced against large scale compressive flows, which act to 
restore the $B - \rho$ relation. There is evidence that in environments such as H I shells, 
which are produced by strong dynamical compression and which show relatively strong 
magnetic fields, turbulence is secondary to flow. This may also be the case in dense 
molecular gas, in which frictional damping of the turbulence is too strong to permit the 
small scale ion-neutral drifts necessary for diffusion. In future work, we intend to 
include full neutral dynamics, which would allow such flows, driven, for example, by 
cooling, or by self gravity. This would also allow an improved realization of the turbulent 
spectrum. Inclusion of a third dimension would permit us to examine the role of field 
aligned flow, and to consider stretching as well as compression of the field. Both can
affect the $B - \rho$ relation.

Finally, we mention some other applications of turbulent diffusion of the magnetic field 
with respect to the gas. It may play an important role in the escape of the large scale 
horizontal field from the Galactic disk. It may also permit the mixing of stellar fields 
with {\textit{in situ}} fields in weakly ionized accretion disks, and jet fields with 
ambient fields in outflows from young stars.

\acknowledgements
We thank N.~C. Brummell for enlightening discussions.
F.~H. is grateful for support by a Feodor-Lynen fellowship of the
Alexan\-der-von-Humboldt Foundation. 
A.~S. acknowledges the support of a Fellowship from the UK
Astrophysical Fluids Facility (UKAFF). The research of J.~E.~G.~D. at Oxford is
funded by the Leverhulme Trust. S. Jansen built the local PC cluster, which
was used, together with the SGI
Origin 2000 machines of the National Center for Supercomputing Applications
at the University of Illinois, for the computations in this paper. This work was
supported by NSF grants AST-0098701 and AST-0328821, and the Graduate School of the 
University of Wisconsin, Madison. The Center for Magnetic Self-Organization in 
Laboratory \& Astrophysical Plasmas is funded by the National Science Foundation.
A minor issue of phrasing was settled with the help of the U.S. Mint.

%
%
\appendix
\section{Measurement of Turbulent Diffusion\label{ss:turbdiff}}

Most discussions of turbulent diffusion are based on quasilinear theory. We provide a
brief review of quasilinear diffusion theory here; for details and a more rigorous 
derivation see e.g. \citet{MOF1978}.

Consider a quantity $q$ which
evolves according to the advection - diffusion equation
\begin{equation}\label{e:adeq}
\left(\ddt+\mbfu\cdot\mbfnabla\right)q=\mbfnabla\cdot\lambda\mbfnabla q
\end{equation}
where $\lambda$ is the microscopic diffusivity.
Assume $q$ can be decomposed into a mean part $\mq$, the ensemble average of $q$,
 and a fluctuating part
$\dq$ with
zero mean, while $\mbfu$ has zero mean and
is isotropic. Averaging equation~(\ref{e:adeq}) gives
\begin{equation}\label{e:madeq}
\f{\partial\mq}{\partial t}=-\langle\mbfu\cdot
\mbfnabla\dq\rangle+\mbfnabla\cdot\lambda\mbfnabla\mq.
\end{equation}
We solve for $\dq$
by subtracting equation~(\ref{e:madeq}) from equation~(\ref{e:adeq}) and discarding the
terms quadratic in the fluctuations. The result is
\begin{equation}\label{e:dadeq}
\f{\partial\dq}{\partial t}=-\mbfu\cdot\mbfnabla\mq+\mbfnabla\cdot\lambda\mbfnabla\dq.
\end{equation}
If the correlation time $\tau$ of the turbulence is short compared to the
diffusion time, the solution of equation~(\ref{e:dadeq}) is approximately
\begin{equation}\label{e:dadeqsol}
\dq\sim-\int^{t}\mbfu\cdot\mbfnabla\mq dt'.
\end{equation}
Substituting equation~(\ref{e:dadeqsol}) into equation~(\ref{e:madeq}) and averaging over
an ensemble gives
\begin{equation}\label{e:madeq2}
\f{\partial\mq}{\partial t}=\mbfnabla\cdot\left(\led +\lambda\right)\mbfnabla
\mq,
\end{equation}
where 
\begin{equation}\label{e:led}
\led\sim u^2\tau
\end{equation}
is the turbulent diffusivity.

Equation~(\ref{e:madeq}) shows that turbulent transport
is the result of the turbulent velocity field $\mbfu$ beating together with
fluctuations in the field to be transported, in this case $q$. With the
assumption that the scales of the mean and fluctuating fields are well separated,
$\dq$ is proportional to the local gradient of $\mq$, and the turbulent flux has
the same form as a diffusive flux. An immediate implication of equation~(\ref{e:madeq}) 
is that if $\mq$ is initially a Fourier mode;
$\mq(\mbfx,0)=\langle\tilde q\rangle\exp{(i\mbfk\cdot\mbfx)}$,
then $\mq$ decays exponentially, with
\begin{equation}\label{e:diffsol}
\mq(\mbfx,t)=\langle\tilde q\rangle\exp{(-\led k^2t+i\mbfk\cdot\mbfx)}.
\end{equation}
In order to investigate whether large scale quantities undergo turbulent diffusion in our
simulations, we integrate equation~(\ref{e:madeq2}) over a domain of area $A$ with
boundary contour $C$. Using Gauss's theorem we derive
\begin{equation}\label{e:mdiffint}
  \f{\partial}{\partial t}\int_{A}\langle q_{k=1}\rangle dxdy=
 \left(\led+\lambda\right)\int_{C}\mbfn\cdot\mbfnabla\langle q_{k=1}\rangle dl,
\end{equation}
where we assume $\led$ and $\lambda$ are spatially constant. The total diffusion coefficient
$D$ is the ratio of the left hand side of equation~(\ref{e:mdiffint}) to the right hand side.

We isolate the large scale fields in our model by setting all modes in the Fourier transform
to zero, except the ones corresponding to $k_{x,y}=\pm 1$. We label the resulting quantity
$q_{k=1}$. We identify the average of $q_{k=1}$ over one GP flow period $\tGP$, 
$\langle q_{k=1}\rangle$, with $\mq$. Since the domain average of $\mq$ is zero, we 
choose $A$ to cover one quarter of the computational domain, centered on 
the maximum of the initial magentic field perturbation. Computing the left and
right hand sides of equation~(\ref{e:mdiffint}) leads to an expression for $D$ 
\begin{equation}\label{e:compeffdiffusion}
D=\f{\partial}{\partial t}\int_A \langle q_{k=1}\rangle dxdy /
  \int_{C}\mbfn\cdot\mbfnabla \langle q_{k=1}\rangle dl.
\end{equation}
%

%
%


\begin{thebibliography}{}

\bibitem[Bourke et al.(2001)]
        {BMR2001}
        Bourke, T.~L., Myers, P.~C., Robinson, G., \& Hyland, A.~R.\ 2001,
        \apj, 554, 916
\bibitem[Brandenburg \& Zweibel(1995)]
        {BRZ1995}
        Brandenburg, A. \& Zweibel, E.~G.\ 1995, \apj, 448, 734
\bibitem[Brumell, Cattaneo \& Tobias(2001)]
        {BCT2001}
        Brummell, N.~H., Cattaneo, F., Tobias, S.~M.\ 2001,
        {\em Fl. Dyn. Res.}, 28, 237
\bibitem[Crutcher(1999)]
        {CRU1999}
        Crutcher, R.~M.\ 1999, \apj, 520, 706
\bibitem[Drazin(1992)]
	{DRA1992}
	Drazin, P.~G.\ 1992, {\em Nonlinear Systems}, pp 140 - 143 (Cambridge: Cambridge
	University Press)
\bibitem[Fatuzzo \& Adams(2002)]
        {FAD2002}
        Fatuzzo, M.~\& Adams, F.~C.\ 2002, \apj, 570, 210
\bibitem[Galloway \& Proctor(1992)]
        {GAP1992}
        Galloway, D.~J. \& Proctor, M.~R.~E.\ 1992, \nat, 356, 691
\bibitem[Goldreich \& Sridhar(1995)]
	{GOS1995}
	Goldreich, P.M. \& Sridhar, S.\ 1995, \apj, 438, 763
\bibitem[Heitsch \& Zweibel(2003)]
        {HEZ2003}
        Heitsch, F. \& Zweibel, E.~G.\ 2003, \apj, 583, 229
\bibitem[Kim \& Diamond(2002)]
        {KID2002}
        Kim, E.-J. \& Diamond, P.~H.\ 2002, \apjl, 578, 113
\bibitem[Kulsrud \& Pearce(1969)]
        {KUP1969}
        Kulsrud, R. \& Pearce, W.~P.\ 1969, \apj, 156, 445
\bibitem[Mac Low et al.(1995)]
        {MNK1995}
        Mac Low, M.-M, Norman, M.~L., K\"onigl, A. \& Wardle, M.\ 1995,
        \apj, 442, 726
\bibitem[Mac Low \& Smith(1997)]
        {MAS1997}
        Mac Low, M.-M \& Smith, M.~D.\ 1997, \apj, 491, 596
\bibitem[Mestel(1985)]
        {MES1985}
        Mestel, L.\ 1985, in Protostars and Planets II,
        ed. D.~C. Black \& M.~S. Matthews
        (Tucson: Univ. Arizona Press), 320
\bibitem[Moffatt(1978)]
	{MOF1978}
	Moffatt, H.K. \ 1978, in {\em Magnetic field generation in electrically conducting
	fluids}, Cambridge Univ. Press, Cambridge
\bibitem[Sarma et al.(2002)]
	{STC2002}
	Sarma, A.P., Troland, T.H., Crutcher, R.M. \& Roberts, D.A.\ 2002, \apj, 580, 928
\bibitem[Soward(1994)]
        {SOW1994}
        Soward, A.~M.\ 1994, in Lectures on Solar and Planetary Dynamos,
        Publications of the Newton Institute, Vol. 2.,
        Cambridge University Press, 181
\bibitem[Strauss(1976)]
	{STR1976}
	Strauss, H.R. \ 1976, Phys. Fluids, 19, 134
\bibitem[Tang \& Xu(2000)]
        {TAX2000}
        Tang, H.~Z. \& Xu, K.\ 2000, J. Comp. Phys., 165, 69
\bibitem[T\'{o}th(1995)]
        {TOT1995}
        T\'{o}th, G.\ 1995, \mnras, 274, 1002
\bibitem[Troland \& Heiles(1986)]
        {TRH1986}
        Troland, T.~H. \& Heiles, C.\ 1986, \apj, 339, 345
\bibitem[Xu(1999)]
        {XUK1999}
        Xu, K.\ 1999, J. Comp. Phys., 153, 334
\bibitem[Zweibel(2002)]
        {ZWE2002}
        Zweibel, E.~G.\ 2002, \apj, 567, 962
\end{thebibliography}
\end{document}